\documentclass[usenatbib]{mn2e}
\usepackage{epsfig,subfigure,amssymb}
\newcommand{\ltaraw}{$\; \buildrel < \over \sim \;$}
\newcommand{\lta}{\lower.5ex\hbox{\ltaraw}}
\newcommand{\gtaraw}{$\; \buildrel > \over \sim \;$}
\newcommand{\gta}{\lower.5ex\hbox{\gtaraw}}

\newcommand{\kms}{{\rm\,km\,s^{-1}}}
\newcommand{\kpc}{{\rm\,kpc\,}}

\newcommand{\msun}{{\rm\,M_\odot}}

\newcommand{\hubunits}{{\rm km\,s^{-1}\,Mpc^{-1}}}

\newcommand{\ffffff}[1]{\mbox{$#1$}}
\newcommand{\scnd}{\mbox{\ffffff{''}\hskip-0.3em.}}
\newcommand{\scmd}{\mbox{\ffffff{''}}}
\newcommand{\ER}{ER~0047--2808}
\newcommand{\apj}{ApJ}
\loadboldmathitalic 
\title[The lens and source of \ER]
{The lens and source of the optical Einstein ring gravitational lens \ER}
\author[R. B. Wayth et al.]
{
Randall B. Wayth$^1$, 
Stephen J. Warren$^2$,
Geraint F. Lewis$^3$, 
\&
Paul C. Hewett$^4$ \\
$^1$School of Physics, University of Melbourne, Victoria 3010, Australia:\tt{rwayth@physics.unimelb.edu.au}\\
$^2$Blackett Laboratory, Imperial College, Prince Consort Rd., London SW7 2BW, U.K.:{\tt s.j.warren@ic.ac.uk} \\
$^3$Institute of Astronomy, School of Physics, University of Sydney, NSW 2006, Australia:\tt{gfl@physics.usyd.edu.au}\\
$^4$Institute of Astronomy, Madingley Rd., Cambridge, CB3 0HA, U.K.:{\tt phewett@ast.cam.ac.uk}}
\date{\today}
\begin{document} 
\maketitle 
\begin{abstract}

We perform a detailed analysis of the optical gravitational lens \ER\
imaged with WFPC2 on the Hubble Space Telescope.  Using software
specifically designed for the analysis of resolved gravitational lens
systems, we focus on how the image alone can constrain the mass
distribution in the lens galaxy.  We find the data are of sufficient
quality to strongly constrain the lens model with no \emph{a priori}
assumptions about the source.  Using a variety of mass models, we find
statistically acceptable results for elliptical isothermal-like models
with an Einstein radius of 1.17\scnd\ An elliptical power-law model
($\Sigma \propto R^{- \beta}$) for the surface mass density favours a
slope slightly steeper than isothermal with $\beta = 1.08 \pm 0.03$.
Other models including a constant M/L, pure NFW halo and
(surprisingly) an isothermal sphere with external shear are ruled out
by the data. We find the galaxy light profile can only be fit with a
S\'ersic plus point source model.  The resulting total M/L$_B$
contained within the images is $4.7 h_{65} \pm0.3$.  In addition,
we find the luminous matter is aligned with the total mass
distribution within a few degrees. This is the first time a resolved optical gravitational lens image has
been quantitatively reproduced using a non-parametric source.

The source, reconstructed by the software, is revealed to have two bright
regions, with an unresolved component inside the caustic and a
resolved component straddling a fold caustic. The angular size of the
entire source is $\sim 0.1$\scmd and its (unlensed) Lyman-$\alpha$
flux is $3 \times 10^{-17}$ erg s$^{-1}$ cm$^{-2}$.

\end{abstract}

\section{Introduction} \label{introduction}

The discovery \citep{1988Natur.333..537H,Langston1989AJ.....97.1283L}
and subsequent modelling
\citep{Kochanek1992ApJ...401..461K,1995ApJ...445..559K} of Einstein
ring gravitational lenses at radio wavelengths raised the prospect of
obtaining detailed quantitative information for both the lensed
sources and the form of the mass distribution in the intervening
lensing galaxies.
\citet{1995ApJ...445..559K} compared mass models for the lensed radio lobe MG1654+134 using \textsc{LensCLEAN} \citep{1992ApJ...401..461K} in image space. It was subsequently shown that \textsc{LensCLEAN}-ing radio data in image space can introduce biases, hence visibility-based \textsc{LensCLEAN} was introduced \citep{1996ApJ...464..556E,2004MNRAS.349....1W}. \citet{2004MNRAS.349...14W} recently used this method to measure $H_0$ in B0218+357.

Recent work at optical/near-IR wavelengths \citep{2000ApJ...535..692K,2001ApJ...547...50K} has concentrated on what can be learnt using infrared images of Einstein rings.
\citet{1992MNRAS.259P..31M} pointed out that there should
exist a much larger number of Einstein rings detectable at optical
wavelengths, and systematic surveys for such systems are now underway
\citep{2000iutd.conf...94H}.
\ER, the first known example of a galaxy
lensing another normal galaxy, was the first Einstein ring to be
discovered at optical wavelengths \citep{Warren1996MN}. The \ER \
system, together with the recently discovered lens 1RXS J113155.4-123155
\citep{2003A&A...406L..43S},
are the only Einstein ring systems with good S/N Hubble Space
Telescope (HST) images at optical wavelengths. Modelling of the
system, exploiting the extended surface brightness distribution of the
lensed source, would provide details of the source morphology and
allow some discrimination between models for the mass distribution in
the intervening lensing galaxy. Compilation of a large sample of
optical Einstein rings could provide information on the luminosity
function and morphologies of star--forming galaxies at redshifts
$z=3-5$ whose (unlensed) magnitudes remain too faint for direct study
using even 8m--class telescopes. Similarly, statistical properties
derived from a large sample of such lensed systems, will provide new
information on the still poorly constrained distribution of mass in
massive early--type galaxies, which form the majority of the deflector
population.

As a first step toward achieving these goals, this paper presents an
analysis of the HST\footnote{Based on observations made with the
NASA/ESA Hubble Space Telescope, obtained at the Space Telescope
Science Institute, which is operated by the Association of
Universities for Research in Astronomy, Inc., under NASA contract NAS
5-26555. These observations are associated with program \# 6560}
imaging observations of \ER.  This system has been studied by
\citet[hereafter `KT03']{2003ApJ...583..606K}, who combined
dynamical information (the stellar velocity dispersion profile), with
gravitational lens information (the angular size of the Einstein
ring), to draw conclusions on the mass profiles of both the visible
and dark matter. The HST image was used simply to determine the total
mass within the ring, and to measure the lens galaxy light profile. In
our complementary analysis, we concentrate on extracting as much
information on the mass profile as possible from the image alone. We
use software based on the \textsc{LensMEM} method \citep[hereafter
`WKN96']{1996ApJ...465...64W} to perform a gravitational--lens
inversion of the complete surface brightness distribution around the
ring. This allows us to draw conclusions on the radial distribution of
the (total) mass, and also produces new insight into the source
morphology.
An advantage of the optical CCD image over radio images is that the counts in the pixels are independent, so that an unambiguous goodness--of--fit statistic may be measured.
To analyse a radio image properly requires the covariance matrix for the image pixels, or to work directly with the visibilities \citep{1996ApJ...464..556E,2004MNRAS.349....1W}.

Section~\ref{observations}
describes the HST WFPC2 observations, the data reduction steps, the
determination of the noise characteristics of the final image of \ER,
and the accurate measurement of the surface brightness profile of the
lensing galaxy.  Section~\ref{gravlens} presents details of the
lens modelling algorithm.  Seven different mass models for the lens are
considered. Four of the mass models provide satisfactory fits to the
data, whereas three are rejected. A discussion of the results, and
comparison with the analysis of KT03, is given in Section 4, before
the principal conclusions are summarised in the final section.

Throughout this paper, we assume a $\Lambda$CDM cosmology with
$\Omega_{\Lambda}=0.7$, $\Omega_{m}=0.3$ and $H_0 = 65 \ \hubunits$
(for consistency with KT03) unless otherwise stated.

\section{Observations and data reduction}\label{observations}

\subsection{Observations and frame combination}

\ER \ was observed with the WFPC2 instrument on HST over four orbits,
on 1999 January 7. The target was placed on chip 4 (pixel size
0.1\arcsec) of the WFPC2 and exposures were made using the F555W
filter, which contains the strong Ly$\alpha$ emission line, at
$z=3.67$, of the lensed source. A dither step was applied between
orbits, on a $2\times2$ grid with step size $N+0.5$ pixels, in order
to improve the sampling. Two equal exposures of $1200\,$s were made
per orbit, with the exception that the first exposure of the first
orbit was only $1100\,$s due to the field acquisition overhead. The
data were processed through the WFPC2 pipeline using the latest
calibration frames.

The first step in combining the frames was to subtract the average
counts in the background from each frame. We then used the {\em
dither} software \citep{2002PASP..114..144F} to identify and eliminate
cosmic rays and then to interlace the data into the half--pixel
grid. In {\em dither} parlance, interlacing corresponds to using a
delta--function {\em drizzle} footprint, {\em pixfrac=0.0}. This was
done to ensure that the data in each pixel remain independent. For the
same reason we did not apply the distortion--coefficient corrections
to linearise the astrometry of the field. The lack of accurate
astrometry over the field is not a concern since we are only
interested in the immediate vicinity of the lensing galaxy.

Cosmic--rays pose a problem since there are only two exposures per
pixel. To identify cosmic--rays we firstly {\em drizzled} the data
using a fat footprint, {\em pixfrac=1.0}, to form a slightly smoothed
combined image. In this way information from adjacent pixels
contributes to the estimate of the counts in a particular interlaced
pixel. The combined image can then be used to identify where the
counts in any individual exposure are abnormally high. Specifically,
the routine {\em driz\_cr} compares the {\em drizzled} image against
the eight individual exposures to identify cosmic rays in each
exposure.  The affected pixels were then masked, an improved average
frame was formed, and further cosmic--rays were identified. In the
region around the galaxy we checked carefully the cosmic--ray
identifications of {\em driz\_cr} in order to ensure the
$\sigma$--rejection level was set at a value close to optimal. The
{\em driz\_cr} routine proved to be highly effective. Once the
cosmic--ray masks for the eight individual exposures had been defined
the data were interlaced, averaging the data pairs, after scaling all
frames to a common exposure time of $1187.5\,$s.

\begin{figure*}
\centerline{
\psfig{figure=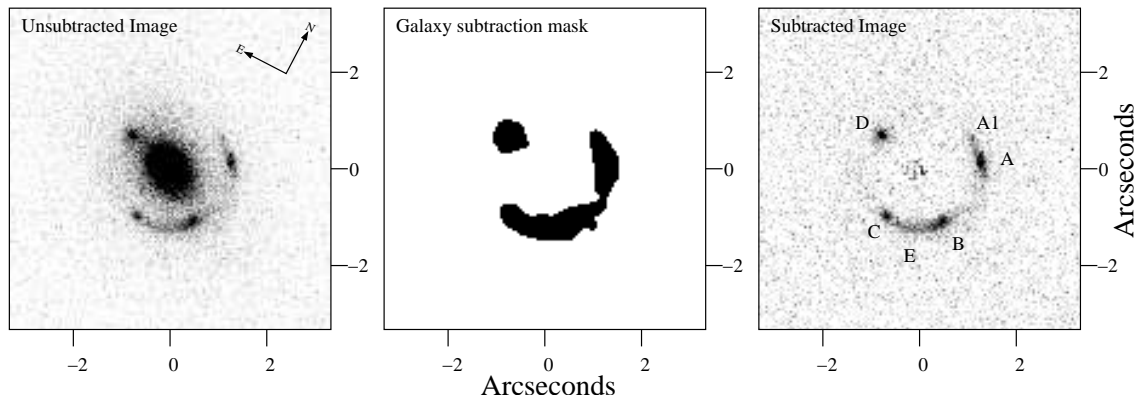,width=150mm,angle=0}
}
\caption{The figure shows a region $6.55\arcsec \times6.55\arcsec$
($131\times131$ pixels) centred on the lens galaxy of \ER; {\it LHS}:
the interlaced final WFPC2 image, {\it centre:} the mask used in the
galaxy fit, {\it RHS}: after subtraction of the galaxy image (as
described in the text). The four bright spots in the image are
labelled A--D. The arc of extended emission between B \& C is labelled
E. The small arc of extended emission north--east of A is labelled A1}
\label{StevesImage}
\end{figure*}

The final combined frame is shown in Fig.~\ref{StevesImage} ({\it
LHS}). The pixels are shown with side 0.05\arcsec, i.e. half the
original pixel size, since this is the interlaced pixel separation.
However, it should be understood that this is for illustration only.
Since the data have been interlaced, the original pixel size, i.e.
0.1\arcsec, has been retained. In other words, Fig.~\ref {StevesImage}
illustrates four frames simultaneously, each offset from each other on
a $2\times2$ grid of side 0.05\arcsec. In all the fits described in
this paper we account for this data format. This is necessary because
of the undersampling of the WFPC2 point spread function (PSF).

The image contains a large number of useful pixels for measuring the
surface brightness distribution of the lens galaxy ($\sim10^4$) and
the Einstein ring ($\sim10^3$). In such cases it is essential to
establish the accuracy to which the uncertainty of the counts in each
pixel is known. This is because the corresponding uncertainty to which
$\chi^2$ may be determined can be as large, or larger, than the width
of the $\chi^2$ distribution itself. For example, if the number of
degrees of freedom in the model fitting ($\simeq$ no. pixels) is
$\sim10^4$, the reduced $\chi^2$ is normally distributed as
$\chi^2_\nu=1.0\pm\sqrt{2/\nu}=1.0\pm0.014$.  For this case, in order
that the uncertainty in $\chi^2$ is not dominated by the uncertainty
in the determination of the noise in each pixel, it would be necessary
to determine the noise to a fractional accuracy substantially better
than $0.007$.

A Poisson estimate of the $1\sigma$ noise--frame was computed,
incorporating read and photon noise, and accounting for elimination of
data due to cosmic--rays, i.e. whether 0, 1, or 2 exposures contribute
to an individual pixel. Standard WFPC2 values of the read--noise,
$5e^-$, and gain, $7e^-/ADU$, were adopted. To check the accuracy of
the estimated uncertainties, the final combined data frame was divided
by the $1\sigma$ noise--frame to form a signal--to--noise ratio (S/N)
frame. The standard deviation of the counts in the background, which
should be unity if the uncertainties are correct, was found to be
$1.06 \pm 0.01$, where the quoted uncertainty in this value was
established from the scatter in measurements from several areas around
the frame. It is unclear whether the small underestimate of the noise
is due to lack of precision in the value of the read noise or the
value of the gain. Therefore, the $1\sigma$ noise--frame was simply
scaled by a factor $1.06$. The accuracy with which this scaling can be
measured, $1\%$, means that the uncertainty in measured values of
$\chi^2$ is $2\%$.

\subsection{Galaxy surface brightness profile}\label{galsub}

To subtract the image of the galaxy, necessary to isolate the image of the
lensed source, we fitted parameterised 2--D models of the galaxy
surface brightness profile convolved with the point spread function,
using a modification of our own software \citep{2001MNRAS.326..759W}
appropriate for {\em drizzled} data. The goodness--of--fit was
calculated by summing $\chi^2$ over the four separate interlaced
frames.  To prevent the presence of the lensed background source from
affecting the model fit to the galaxy, we masked out pixels in the
vicinity of the Einstein ring. This was achieved in an iterative
fashion, firstly subtracting the galaxy fit, then smoothing and
thresholding the residuals to create a first mask, then refitting, and
then refining the mask, etc. The final mask contains 842 pixels and is
shown in Fig.~\ref {StevesImage} ({\it centre}).

Fitting the surface brightness profiles of early--type galaxies, which
are cuspy at the centre, requires care, particularly where the data
are undersampled. To generate a model image for each interlaced frame,
the model galaxy profile was firstly integrated over sub--pixels of
size $0.1\arcsec/3$. The PSF was computed using the {\em tinytim}
software \citep{1993adass...2..536K,1995adass...4..349K}, integrated
over the same sub--pixel size. The model image was then convolved with
the model PSF. The HST PSF is undersampled by WFPC2 at the wavelength
of our observations and the sub--pixelation is necessary to correctly
reproduce the actual image that would be seen by the detector. After
rebinning this convolved image to the full $0.1\arcsec$ pixel size,
the image was further convolved with the WFPC2 pixel scattering
function, as the final step in forming a model image. The pixel
scattering function accounts for the diffusion of electrons from the
collecting pixel into nearby pixels. We used the $3\times3$ kernel
specified in the WFPC2 Instrument Handbook \citep{HST_WFPC_Handbook}.
For this kernel, a fraction 0.365 of the collected electrons diffuse
out of the central pixel, principally into the four nearest pixels. So
the scattering function broadens the PSF significantly. Because of the
cuspy galaxy light profile, the kernel is an essential element in the
model fitting. However, the kernel was determined from pre--flight
measurements, and it has not been remeasured in orbit. Therefore,
somewhat surprisingly, it is a significant source of uncertainty (see
below).

In the following we use $X, Y$ for the coordinates on the CCD, and $x,
y$ for, respectively, the major and minor principal axes of any
ellipse. The radius $R$ is defined by $R^2 = x^2q +
y^2/q$, where $q=b/a$ is the ellipse axis ratio. The galaxy
model parameters are the $X, Y$ position of the centre, the axis ratio
$q$, and the angle of orientation $\phi$, plus the parameters of the
function defining the surface brightness profile. A de Vaucouleurs
profile \citep{1948AnAp...11..247D} did not provide a satisfactory
fit, so a S\'ersic model \citep{1968adga.book.....S}, where the
surface brightness as a function of $R$ is $S(R)=S_{1/2}
\exp\{-B(n)\lbrack(R/R_{1/2})^{1/n}-1\rbrack\}$, was tried. The
parameter $n$ quantifies the shape of the profile: the values $n=0.5$,
$n=1$, and $n=4$ correspond to the Gaussian, exponential, and de
Vaucouleurs profiles. Profiles with larger $n$ are more sharply
peaked.  For cuspy galaxies, $n>3$, special care is needed in
integrating the S\'ersic function over sub--pixels near the
centre. $B(n)$ is a constant for particular $n$, and we used the
series asymptotic solution for $B(n)$ provided by
\citet{1999A&A...352..447C}.

With the S\'ersic function we obtained fits that were globally
satisfactory, in terms of $\chi^2$. Nevertheless the fit over the
central $0.5\times0.5 \arcsec$ was bad. We also found that the fitted
value of $n$ depended sensitively on the size of the box, increasing
as the box size diminished. This could be because either the S\'ersic
profile is not cuspy enough for this galaxy, or the pixel scattering
kernel used is too broad. To investigate the latter we modified the
kernel, reducing the scattered fraction from 0.365 to 0.2, but the
central fit was still unsatisfactory. We therefore retained the
standard kernel, and added a point source at the centre of the model
surface brightness profile. The full model has 8 free parameters; $X,
Y, q, \phi, S_{1/2}, R_{1/2}, n$, and $p$, the counts in the
point source.  Here it should be understood that $R_{1/2}$ is the
half--light radius of the S\'ersic component of the light profile, and
not the radius containing half the total galaxy light. The final fit
was made to a square--region, $131\times131$ {\em drizzled} pixels,
i.e.  $6.55\arcsec\times6.55\arcsec$ in extent, centred on the galaxy,
the entire region reproduced in Fig.~\ref {StevesImage}. The fit is
satisfactory, both globally, and in the central region. 

The image after subtraction of the galaxy profile is shown in
Fig. \ref{StevesImage} ({\it RHS}). We have designated the main
components of the lensed image in a similar fashion to KT03, with the
four main bright spots labelled A, B, C, \& D clockwise from the NW
component. We add the labels E for the arc of extended emission
between images B and C, and A1 for the extended emission north of
A. The increased Poisson noise visible toward the centre of the
subtracted image is a consequence of the steep luminosity
profile. Because of this, a demagnified central (fifth) image of the
source would not be detectable with these data.

The reduced $\chi^2$ of the S\'ersic + point source fit is
$\chi^2_{\nu}=0.96\pm0.02$, whereas the expected value is
$1.000\pm0.011$ for $\nu=16311$. These two values are consistent, and
both the fit and the estimated uncertainties are therefore deemed
satisfactory. (Note that in this case the accuracy to which
$\chi^2_{\nu}$ can be measured ($2\%$, \S2.1) is worse than the
standard deviation of the $\chi^2_{\nu}$ distribution for the given
number of degrees of freedom.)

The parameters of the fit are provided in Table
\ref{tab:phot_fits}. Magnitudes are on the VEGAMAG system for the
F555W filter, and use the zero point 22.538, for chip 4
\citep{HST_WFPC_DataHandbook}.  The orientation of chip 4 was taken
from the image header. The PA of the Y axis of chip 4 was at
$27.44^{\circ}$ for these observations. To calculate the PA,
hereafter $\theta_0$, this angle was added to $\phi$, the measured
orientation of the galaxy image on the chip.

For the best fit $n$, we have $B(n)=5.89$.  For this model, the galaxy
total magnitude (S\'ersic + point source) is $V_{555}=20.61$. KT03
obtained the same value for the total magnitude, fitting a de
Vaucouleurs profile, following a similar procedure to that described
here, but which differs in detail.  However, in common with other
authors \citep[e.g.][]{2001AJ....121.2358B}, we caution that
meaningful total magnitudes cannot be measured for early--type
galaxies. This is made clear when we compare the best--fit S\'ersic
only model, to the S\'ersic + point source fit. In terms of aperture
magnitude the two profiles agree to within a few percent at all radii
over the fitted region, except at the very centre. However, the
best--fit S\'ersic profile (which has $n=6.9$) has very extended
low--level wings beyond the fitted region, and a total magnitude
$40\%$ brighter. This shows that the models are only useful over the
region of the fit. For similar reasons it is not meaningful to compare
half--light radii of fits of different models.

The uncertainties quoted in the table are the random errors, computed
by inverting the $\chi^2$ curvature matrix. The actual uncertainties
in some of the parameters are larger, and are dominated by the
uncertainty of the pixel scattering function. For example, for the
sharper kernel considered above, where the scattered fraction is 0.2,
the parameter $n$ increases to 3.7, and the brightness of the point
source reduces to $V_{555}=24.3$. Although the values of the
parameters of the fits for the different kernels are quite different,
again the profiles themselves are very similar over the region of the
fit, except at the very centre.

\begin{table}
\begin{tabular}{llr}

parameter & unit & value \\ \hline
$R_{1/2}$        & $\arcsec$                   & $1.09\pm0.03$      \\
$S_{1/2}$   & $V_{555}$ mag./sq.$\arcsec$ & $24.11\pm0.05$     \\
$q=b/a$          &                             & $0.69\pm0.01$      \\
$\phi$           &          deg.               & $35.0\pm0.7\:\:$   \\
$\theta_0$       &     deg. E of N             & $62.4\pm0.7\:\:$   \\
n                &                             & $3.11\pm0.09$      \\
p                & $V_{555}$ mag.              & $23.80\pm0.04$     \\ \hline
$V_{555}(tot)$   &                             &  20.61 \\
$M_B(tot) - 5\log h_{65}$ &                    & $-22.22$ \\
\end{tabular}
\caption{Photometric parameters for the lens galaxy, measured assuming
the standard WFPC2 pixel scattering kernel.} 
\label{tab:phot_fits}
\end{table}

We find satisfactory agreement between the S\'ersic + point source
profile and our previous fit of a de Vaucouleurs profile to a ground
based image of the galaxy \citep{Warren1996MN}. The previous measured
values of $q$ and PA are consistent with the new values. The old fit
gave an integrated brightness $V=21.12$ within the measured effective
radius $R_e=1.39\arcsec$. For the S\'ersic + point source model fit to
the HST image, within the same radius we find $V_{555}=21.17$, in
excellent agreement.

To summarise, using an 8--parameter S\'ersic + point source model,
convolved with the HST PSF, we obtained a good fit to the galaxy
surface brightness profile, at all radii over the $6.55\arcsec
\times6.55\arcsec$ box.  The measured values of the parameters are
somewhat sensitive to the adopted pixel scattering
kernel. Nevertheless, over the fitted region, the deconvolved surface
brightness profiles determined are not sensitive to this function. The
galaxy model was subtracted, to leave the Einstein--ring image to be
modelled.

\subsection{Absolute magnitude}

The relation between absolute magnitude in the $B$ band, $M_B$, and
$V_{555}$ is: 
$$M_B=V_{555}-5\log_{10}(d_L/10)-k_{555}+(B-V_{555})_0-g_{555},$$ where
$d_L$ is the bolometric luminosity distance in pc, $k_{555}$ is the
$k-$correction for the $V_{555}$ filter, $(B-V_{555})_0$ is the galaxy
colour in the restframe, and $g_{555}$ is the Galactic extinction for
this galaxy at the wavelength of the $V_{555}$ filter. The combined
term $t=-k_{555}+(B-V_{555})_0$ depends on the spectrum of the galaxy
(corrected for Galactic reddening). To determine $t$, KT03 computed
synthetic spectra that match the measured colour of the galaxy
$V_{555}-I_{814}=1.94\pm0.12$, finding $t=-0.439$ (properly the colour
should be corrected for Galactic reddening, but the correction is very
small compared to the colour photometric error). Note that the colour
of an old elliptical at this redshift would be $V_{555}-I_{814}\sim2.4$.

We have performed similar calculations, using the BC03 population
synthesis code \citep{2003MNRAS.344.1000B}, for galaxies of solar
metallicity, and of specified formation redshift $z_f$, with an
exponentially declining star formation rate, of timescale
$\tau$. Combinations $z_f=2$, $\tau=1$Gyr, and $z_f=4$, $\tau=1.5$Gyr,
reproduce the measured colour, and yield similar values of $t$ to that
computed by KT03. Therefore we have adopted their value for $t$, as
well as their value $g_{555}=0.052$. The final result is
$M_B=V_{555}-42.834 + 5 \log h_{65}$, used to compute the value of $M_B(tot)$ quoted
in Table \ref{tab:phot_fits}.

The BC03 model spectra provide an additional useful quantity, the
restframe $B-$band luminosity evolution between $z=0.485$ and 0. The
results for the two models are $\Delta m=0.9$mag.

\section{Gravitational Lens Inversion}\label{gravlens}

The HST image, Fig.~\ref {StevesImage}, shows four bright peaks. The
lensing analysis by KT03 was limited to fitting the positions of the
four peaks in the ring, on the assumption that these are all images of
the same source. This is similar to the procedure followed in
analysing images of gravitationally--lensed quasars, and is usually
satisfactory for accurate measurement of the total mass within the
ring.  Nevertheless, gravitational lenses like \ER, for which the
source is resolved, offer the prospect of measurement of the profile
of the mass distribution in the lens, as well as the light profile of
the source.  Hitherto, the small number of radio Einstein rings have
provided the only systems with deflectors of galaxy mass, possessing
such an extended distribution of surface brightness. The detailed
structure revealed by the WFPC2 imaging makes \ER\ the first system
discovered at optical wavelengths for which a similar analysis is
possible. As detailed below, we have discovered that the four peaks
are in fact due to two distinct sources.

\subsection{Inversion algorithm}

Presented with an image such as Fig.~\ref {StevesImage}, the problem
is to determine simultaneously the source light profile and the lens
mass profile. For a review of the different published solutions to
this problem, the reader is referred to the Introduction section in
\citet{2003ApJ...590..673W}.  A non--parametric solution, where the
source light profile is pixelised, is preferred to a parametric
solution, because the light profiles of high--redshift galaxies are
typically complex and there is no clear way to choose an appropriate
parameterisation. On the other hand the surface mass distribution in
the elliptical galaxy lens is expected to be much simpler, and may be
parameterised in the first instance (i.e. until a poor fit indicates
the model needs refining e.g. to incorporate additional mass or shear components). A proper
solution must account for the smearing of the image by the
PSF. Because of this deconvolution step, some form of regularisation
will usually be required. The \textsc{LensMEM} algorithm described by WKN96
satisfies all these requirements, and is the method we have used
here. We follow closely the methods set out in WKN96 with a few
significant improvements. Here we provide only an outline of the
method, and detail any differences.

The goal of the process is to find a source and lens model which
provide an acceptable fit to the data. The quality of the fit to the
data is quantified by the $\chi^2$ statistic. The merit function that
is minimised is $C = \chi^2 - \lambda S$, where $S$, the regularising
term, is the entropy in the source plane.
The term $\lambda$ is a
weight which determines the relative importance of the $\chi^2$ and
entropy terms to the solution.
The adjustment of $\lambda$ is explained below.

Using $s_i$ for the counts in source pixel $i$, the usual expression
for entropy is $S=-\sum_{i} s_i\ln{s_i}$. We have used the the
modified term $S= -\sum_{i} s_i [\ln(s_i/A)-1]$
\citep{1984MNRAS.211..111S}. With this expression $A$ defines a default
source pixel value. We found the solution to be
insensitive to this parameter, for appropriately small values,
below the noise level in the image.

The WKN96 algorithm proceeds in three nested cycles. The innermost
cycle determines the solution for the source light distribution for a
fixed set of lens mass parameters, and for fixed $\lambda$. The middle
cycle adjusts $\lambda$.
The outer cycle adjusts the lens mass parameters and calculates the mapping matrix.
For a fixed mass model, a mapping matrix is generated by
dividing each pixel in the image into two triangles which are
projected back into the source plane. The mapping matrix records the
fraction of each source pixel from which a given image pixel maps. Once
computed, the mapping matrix has convenient properties which allow
projection between source and image plane (in either direction,
correctly conserving surface brightness) and the calculation of
critical lines and caustics. The model image is formed by projecting
the source through the mapping matrix, followed by convolution with
the PSF.
If $\lambda$ is large, the minimum$-C$ solution will have a smooth source, but the quality of
the fit, as quantified by $\chi^2$, may be unsatisfactory. The
$\chi^2$ of the fit may be improved by reducing $\lambda$. The curve
of $\chi^2$ (at the minimum$-C$ solution) against $\lambda$ is
monotonic.  The goal is to have the smoothest solution for the source
that gives a `satisfactory' fit, defined by a target value of
$\chi^2$. The middle cycle adjusts $\lambda$ searching for the
target $\chi^2$ point on the curve for a fixed mass model.

WKN96 used a conjugate gradient technique with search direction $\nabla C = \nabla
\chi^2 - \lambda \nabla S$ where $\nabla \chi^2$ is the gradient of
the $\chi^2$ projected back into the source plane. For fixed
$\lambda$, each iteration in the inner cycle applies a scaled $\nabla
C$ to the current source model such that $C$ is minimised. The inner
cycle stops when no more improvement can be made.  There are two
problems with WKN96's implementation.  Firstly, the magnitude of
$\nabla S$ can be very large for individual pixels which happen to be
small positive numbers due to the gradient of the entropy
$\partial S / \partial s_i = \ln A - \ln s_i$.
Secondly, adding the middle cycle, to adjust $\lambda$, is inefficient.
\citet{1984MNRAS.211..111S}
identified these problems (and more) and created an elegant algorithm
to find solutions which incorporate adjusting $\lambda$ as part of the
inner cycle. In this way there are only two cycles. For a fixed mass
model, the algorithm adjusts $\lambda$ and the source light profile at
the same time, to achieve the maximum--entropy solution subject to the
constraint that $\chi^2$ equals the target value. The mass model is
then adjusted to give the maximum of these maximum entropy solutions. The
method achieves the same result as before, but views the problem
differently. The term $\lambda$ is now seen as a Lagrange multiplier
in a constrained minimisation problem, rather than as a weight on the
regularisation term.  We have used the algorithm from
\citet{1984MNRAS.211..111S} in our implementation,
and have found it superior to the conjugate gradient technique
used in WKN96.

\subsection{Mass Models}
\begin{table*}
\begin{tabular}{l|l|l}
Model & Definition & Free Parameters \\ \hline
1. PIEP & $\psi(R) = bR$ & $b, q, \theta_0$ \\ 
2. SIS+$\gamma$ & $\psi(R,\theta) = bR + \frac{\gamma}{2}R^2\cos 2(\theta - \theta_0)$ & $b, \theta_0, \gamma$ ($q=1$)\\
3. SIE & $\kappa(R) = (b/2)R^{-1}$ & $b, q, \theta_0$ \\
4. SPEMD & $\kappa(R) = (b/2)R^{-\beta} $ & $b, q, \theta_0, \beta$ \\
5. Constant M/L & $\kappa(R) = \kappa_{1/2} S(R)$ & $\kappa_{1/2}$ \\
6a. NFW & $\rho(r)=\frac{\rho_c}{(r/r_s)(1+r/r_s)^2}$ & $\rho_c, q, \theta_0, r_s=25\kpc$ \\
6b. NFW & $\rho(r)=\frac{\rho_c}{(r/r_s)(1+r/r_s)^2}$ & $\rho_c, q, \theta_0, r_s=5\kpc$
\end{tabular}
\caption{Summary of lens models used in this paper. The models are
defined by either their 2-dimensional lens potential, $\psi(R)$, their
dimensionless surface mass density, $\kappa(R)$, or their volume density, $\rho(r)$.
The radial coordinate is defined as $r = [x^2(1-\epsilon) + y^2(1+\epsilon)]^{1/2}$
or $R = (x^2q + y^2/q)^{1/2}$. $S(R)$ is the S\'ersic function described in \S \ref{galsub}
We characterise the isothermal-like models by their Einstein radius $b$.}
\label{tablensmodels}
\end{table*}

A primary goal of the study is to achieve some discrimination between
the fairly extensive range of mass/potential models that have been
found to reproduce image configurations in gravitationally lensed
systems that possess fewer observational constraints
\citep[e.g.][]{1995ApJ...445..559K}. We have tested six simple
elliptical models. The definitions of the models are
provided in Table \ref{tablensmodels}.
The models are defined in terms of
the radial profile of either the potential, $\psi$, the dimensionless surface mass
density, $\kappa$, or the volume density, $\rho$. We fixed the centre of each model to
be the centre of the light distribution.
The six models are:
1. the pseudo-isothermal elliptical potential (PIEP) \citep{1987ApJ...321..658B},
2. the singular isothermal sphere in an external shear field (SIS+$\gamma$),
3. the singular isothermal ellipsoid (SIE) \citep{1993ApJ...417..450K,1994A&A...284..285K},
4. the power-law ellipsoid (SPEMD) \citep{1998ApJ...502..531B} (of which the SIE
is a special case),
5. the constant mass--to--light ratio (M/L) model, defined by the S\'ersic+point--source profile fit described in Section \ref{galsub}, and
6. the NFW model \citep[hereafter NFW96]{1996ApJ...462..563N}.
For each model the deviation from circular symmetry is parameterised
by the axis ratio $q=b/a$
or the external shear $\gamma$ (for the SIS+$\gamma$).
The profiles are referred to the
ellipse coordinate $R = (x^2q + y^2/q)^{1/2}$, where $x$ and $y$ are
the major and minor principal axes with PA $\theta_0$.
The mass scale $b$ is normalised in angular units to the usual
definition for the Einstein radius of a SIS:
$b = 4\pi (\sigma^2/c^2) (D_{ds}/D_{s})$.
Note that for the SPEMD,
$b$ is the Einstein radius only when $\beta=1$, otherwise it
should be viewed as a dimensionless scale factor.
The Constant M/L model mass scale has units of $\Sigma_{crit}$, while
the NFW model has units of $\Sigma_{crit}/r_s^2$ where $\Sigma_{crit} = (c^2/4 \pi G)(D_s / D_d D_{ds}) = 1690 h_{65}^{-1} \msun$ pc$^{-2}$
is the critical density of the lens.

For the NFW parameterisation we fitted two models, with different
values of the scale radius $r_s$.  The measured stellar central
velocity dispersion for the lens is \mbox{$229 \pm 15$ km s$^{-1}$} (KT03). By
comparison with the $n-$body simulations of \citet{2001MNRAS.321..559B}, for
the galaxy redshift $z=0.485$, the corresponding virial mass is
$\sim10^{13}\msun$, and an appropriate value of the scale length is
$r_s\sim50$kpc. However the effect of the baryons would be to steepen
the mass profile, and a smaller scale length would be appropriate. The
scale lengths we chose were 25\kpc (model 6a) and 5\kpc (model
6b). The model with the smaller scale radius would be expected to have
similar lensing characteristics to the SIE model, since the NFW
density profile is approximately $\propto r^{-2}$ around the scale radius.

Deflection angles can be simply computed from the potentials for the
PIEP and SIS+$\gamma$ models, while the calculation is analytic for
the SIE model. We use the FASTELL code \citep{1998ApJ...502..531B} to
compute deflection angles for the SPEMD model. The deflection angle for
the constant M/L model must be computed numerically, for which we
followed the prescription of \citet{2001KeetonB}. For the NFW model we
used the approximation introduced by \citet{2002A&A...390..821G} to
compute deflection angles.

\subsection{Source--plane pixelisation}
We chose a square source plane, offset from the centre of the
galaxy. The size and position of the source plane were finalised after
preliminary modelling provided an indication of the approximate
location and extent of the source. The source plane should be as small
as reasonably possible. If the source plane is made too large, most of
the pixels will map to sky in the image plane, and this reduces the
power of the $\chi^2$ statistic to discriminate between different mass
models.  We used a $10\times10$ grid of pixels.
The source pixel size was chosen such that the size of the caustic (in pixels) for different mass models was approximately the same.
The pixel size was chosen as a reasonable match to critical sampling of the image-plane resolution element transformed to the source plane. This criterion ensures that the full information content of the image is used in the source reconstruction, while minimising the degree of regularisation required [these issues are discussed more fully in \citet{2003ApJ...590..673W} and \citet{2004DyeWarren}].
The source pixel size for all models was within the range $0.05\arcsec/1.5$ and $0.05\arcsec/3$.

Mapping the entire source plane to the image plane defines the minimum
extent of the region over which $\chi^2$ should be measured. During
the minimisation process, as the mass parameters vary, the size of
this region varies (since the mapping depends on the mass
parameters). This can mean that the surface of the merit function is
not smooth, since for small changes in a mass parameter the number of
image pixels varies, as a pixel is included or excluded. For this
reason we defined a fixed region, or `mask', in the image plane of
822 pixels over which $\chi^2$ is measured. The mask is chosen to be
slightly larger than the image of the source plane in order to allow for
size variations in the projected image of the source plane during minimisation.

\subsection{Discriminating between models}
Each model has 3 (or 4 for the SPEMD, 1 for the Const M/L) parameters for the mass model plus
100 for the source plane pixels. With the fixed 822 pixel mask, there
are then 719 degrees of freedom in the model and the acceptable range of
$\chi^2$ is $719 \pm 38$.

Models which do not produce a satisfactory fit in terms of $\chi^2$
are ruled out. However a more stringent comparison is possible with
the F--test. The principle of the F--test is as follows; for a model
that is linear in the parameters, the significance of the improvement
in $\chi^2$ achieved by the addition of $N$ extra parameters may be
assessed by comparison of $\Delta\chi^2$ against the $\chi^2$
distribution for $N$ degrees of freedom. This must be applied here
with a degree of caution, since we are not adding parameters to linear
models, but comparing different non--linear models.

There is an additional complication associated with the application of
regularisation in the inversion via the entropy constraint.
Regularisation smooths the source.
This acts to reduce the effective number of parameters in the fit, and increases the effective
number of degrees of freedom, by an amount which increases with the
degree of regularisation. Unfortunately, there is no way to quantify
this increase of degrees of freedom. However, the entropy constraint is similar
to a `zeroth-order' regularisation constraint in that it is the sum
of contributions from individual pixels and has no inter-pixel dependence.
Likewise the gradient of the entropy
$\partial S / \partial s_i$ depends only on an individual pixel's value.
Hence, the covariance between pixels is generated only through the overall
contribution of $\lambda S$ to $C$.
The use of a small source also helps to minimise the (unknown) increase by
only using pixels which are required by the data.
In addition, rather than follow the usual procedure of searching for the
smoothest source consistent with the target $\chi^2$ (which one would do if
particularly interested in the nature of the source when the mass model is known),
we instead seek the best fit achievable for each mass model.
We implement this simply by setting the target $\chi^2$ to be 680 for all models
and note that no mass models were able to reach this lower limit on the range of
acceptable $\chi^2$. Hence, the procedure is effectively a non-negative min$-\chi^2$ solution.
This procedure again minimises the increase of degrees of freedom.
(For these solutions discussed in \S \ref{sec:results}, the entropy term in the merit function, $\lambda S$, is a factor at least 70 times smaller than the $\chi^2$ term.)
Furthermore, our scaling of the size of the source region to the size of the caustic
for each model, ensures that each model is treated in a comparable
way. In this way any (small) increase in the number of degrees of
freedom will be quite similar for each model, and therefore will have
minimal effect when comparing models. Accordingly we ignore this
uncertainty, but demand a high level of significance in ruling out
models.

\subsection{Errors on parameters}

If regularisation is applied, the only correct way to measure the
errors is through Monte Carlo methods, applying random realisations of
noise to the data, and measuring the spread in the parameters. However
 the degree of regularisation applied here is minimal because the fit
amounts to a non--negative min$-\chi^2$ solution, in which case the
errors may be estimated by inverting the Hessian matrix for the
$\chi^2$ surface at the minimum. We verified that the Hessian matrix
and Monte Carlo methods provided very similar answers for one model,
and then used the Hessian matrix for all other models.

\section{Results}\label{sec:results}

\subsection{The lens galaxy}\label{sec:lensgal}

\begin{figure*}
\subfigure{\includegraphics[scale=0.80]{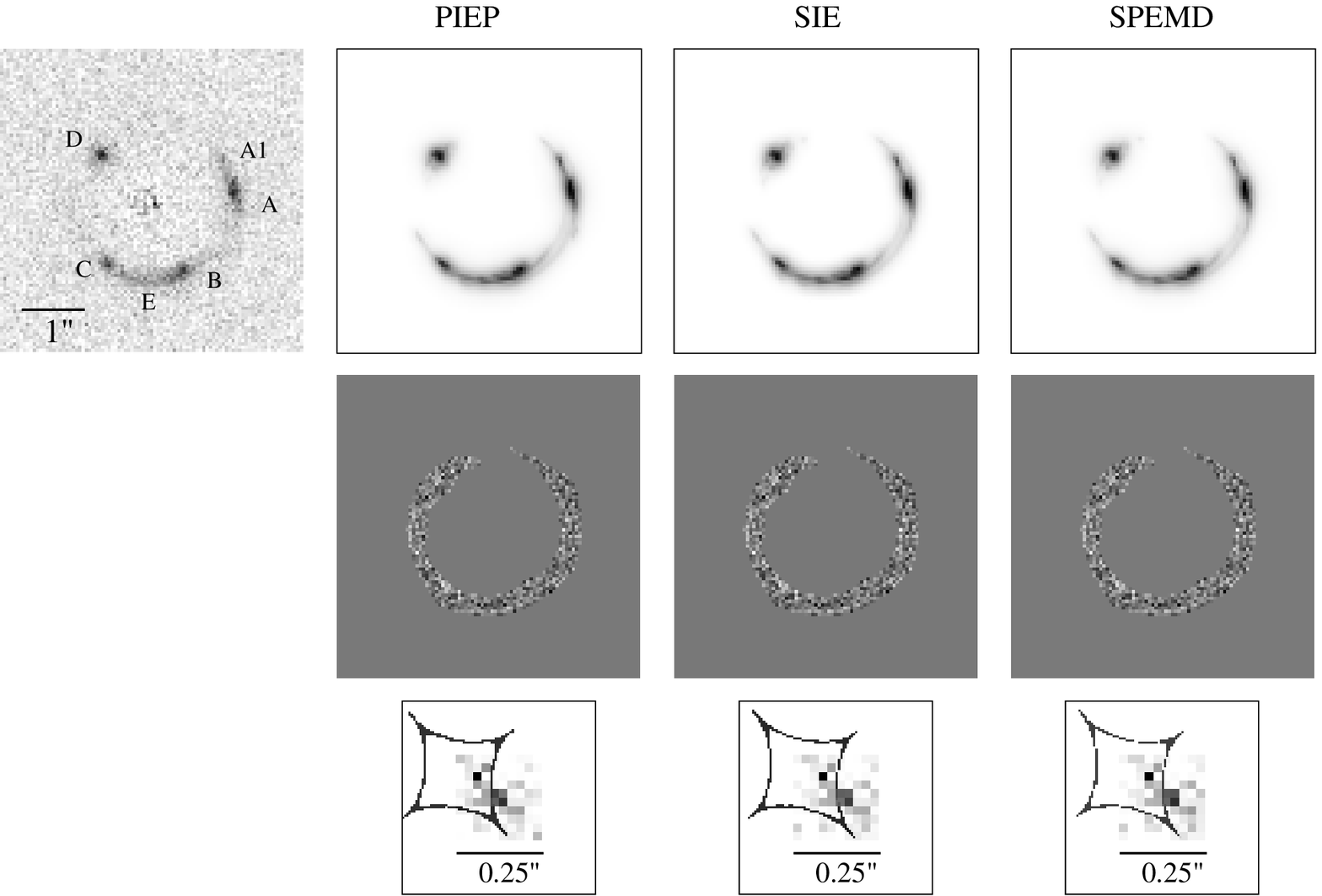}}
\subfigure{\includegraphics[scale=0.80]{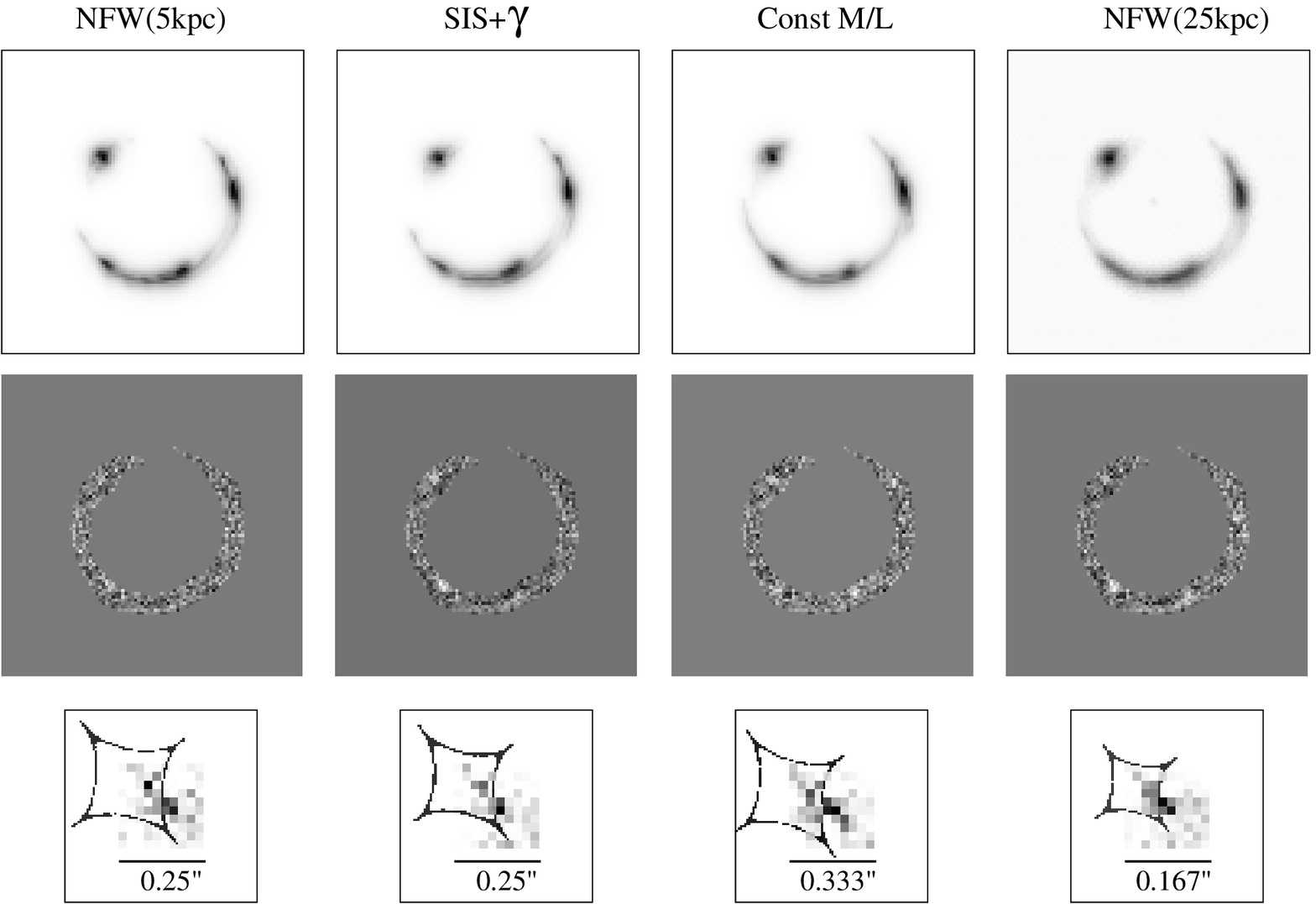}}
\caption{Data and best fitting images for each of the lens models are
shown in the upper panels. The residual between the data and model is shown in the centre panels.
The corresponding sources, together with the model caustics, are shown in the lower panels.}
\label{fig:Modelimgs}
\end{figure*}

The results of the modelling procedure are provided in Table \ref{bigtable}, and
illustrated in Fig. \ref{fig:Modelimgs}.
Table \ref{bigtable} lists in successive columns: 1. the name
of the lens model, 2. the number of degrees of freedom and $1\sigma$
uncertainty, 3. the $\chi^2$ of the best fit, 4. the ratio of the image pixel
size to the source pixel size, 5. the area of the
caustic, and 6. the relevant best--fit parameters and their errors.

Notwithstanding the above caveats on the comparison of models, the
interpretation of these results is relatively straightforward. Two
models, the SIS+$\gamma$ and NFW(25\kpc) models, are strongly rejected
on the basis of $\chi^2$ alone. Four of the models, the PIEP, SIE, SPEMD,
and NFW(5\kpc) models, provide acceptable fits, with very similar
values of $\chi^2$.
This is not too surprising, since these four
models all have very similar mass profiles over the radii of
interest.
The axis ratio $q_{PIEP}$ can be calculated for equivalent \emph{isodensity}
contours of the PIEP model (for small ellipticities) using equation 3.3.2
of \citet{1993ApJ...417..450K}. This yields $q_{PIEP}= 0.77$ in excellent agreement
with the SIE/SPEMD. This confirms the expected result that the models are
equivalent for large $q$.
Finally the constant M/L model lies between the two
extremes, producing a $\chi^2$ which is rejected with 98\% confidence.

Applying the F-test to our models, extends these results. For example
comparing the SIE model and the constant M/L model, we have
$\Delta\chi^2=86$ for a change in number of degrees of freedom of only
2. The significance of this is so great that we may safely rule out
the constant M/L model. However the most interesting case is
comparison of the SPEMD and SIE models. Here the only change is between
fixing the power law exponent to the value $\beta=1$, and allowing it
to be a free parameter. The best fit value for the SPEMD model is
$\beta=1.08\pm0.03$ i.e. a small but significant change. The change
$\Delta\chi^2=4.7$ for $\Delta N=1$ is significant at the $97\%$
level.

Thus, the overall mass distribution in the lens galaxy favours
a mass power law which is slightly steeper than isothermal around the
Einstein radius.  This is important because $H_0$ measurements based
on lensing time delays are sensitive to the mass power law ($\Sigma
\propto R^{-\beta}$) at the radius of the images, with larger values of $\beta$ (i.e. more centrally
concentrated mass) leading to larger actual time delays. (This would then be interpreted as smaller value of $H_0$ using an isothermal model).
Most importantly, we found no strong degeneracy between mass power law
$(\beta)$, mass scale $(\Sigma_0)$ and axis ratio $(q)$ for the SPEMD
model-- a problem frequently encountered when modelling quasar lenses
\citep[e.g.][]{1994AJ....108.1156W}.

The mass enclosed within the critical line is $3.05 h_{65}^{-1}
\times 10^{11} \msun$ for all successful mass models.
Using the photometric properties of the lens galaxy
from Table \ref{tab:phot_fits}, we calculate the total projected
$M/L_B$ inside the Einstein radius to be $4.7 h_{65} \pm 0.3$ $M_{\odot}/L_{B\odot}$. This value is
smaller than that calculated by KT03 ($5.4 \pm 0.8$), however KT03
used a larger Einstein radius (1.34\scmd) and the presence of dark
matter would make the total $M/L_B$ increase with radius.
We also note that the constant M/L model, if correct, would require the
stellar $M_*/L_B$ to be $6.8 h_{65}$.

Finally, we can infer some qualitative properties of the galaxy's dark
halo in the region of the Einstein radius.
The position angle of the mass distribution is
almost identical for all successful models and is within a few degrees
of the measured position angle for the lens galaxy. The axis ratio of
the overall mass profile (q=0.77) is rounder than the axis ratio of
the visible galaxy (q=0.69). This indicates that the halo must be
substantially rounder than the visible galaxy.

\subsection{The source}\label{sec:source}

The inferred morphology of the (unlensed) source, Fig. \ref{fig:Modelimgs}, is particularly
interesting. For all the models which provided satisfactory fits the
reconstructed source is double, with the smaller component lying inside a fold
caustic, and the larger, equally bright, component lying on top of the
same fold caustic. This vindicates the use of a non--parametric model
for the source. It would be very difficult to determine this solution
for the source with a parametric model.

\begin{figure}
\centerline{
\psfig{figure=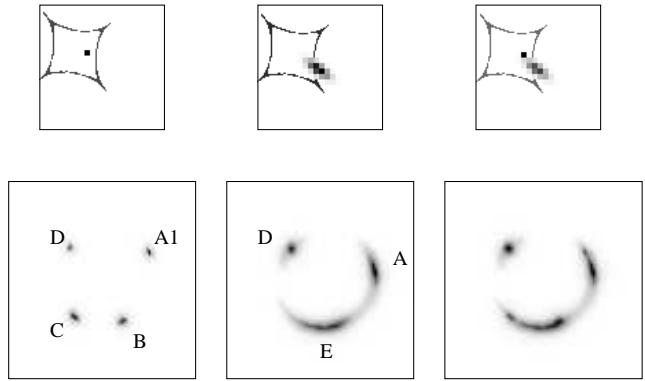,width=85mm}
}
\caption{The two components used as the model source (top row)
and the corresponding images (bottom row). The linear size of the
source plane shown here is 8 times its true size relative to the image plane.}
\label{fig:Modelsrcs}
\end{figure}

Guided by these results, we created a simple two-component source
model, lensing each component separately, and together, through the
PIEP lens model, to create the images shown in Fig.
\ref{fig:Modelsrcs}. The lower left panel shows the image of the inner
source component, modelled as a single bright pixel. The middle lower
panel is the image of the outer source component, modelled as an
elliptical Gaussian. The combination image is shown in the lower right
panel. Comparing against Fig. \ref{StevesImage}, this model shows that features B, C
and the brighter spot in A1 are unresolved images of a single source.
Features E, A and most of A1 are the image of the extended object
which crosses the caustic. Feature D is the combined, barely resolved,
image of both components. In their modelling KT03 assumed that the
images A, B, C, D are the four images of a single source
component. However they were unable to obtain a good fit with their
model. This further illustrates the importance of using a suitably
detailed source model, also illustrated in \citet{2003ApJ...590..673W}

In contrast with the models which provide good fits, the source
reconstructions for the SIS+$\gamma$, NFW(25kpc), and constant M/L
models consist of several fainter blobs or are smeared out and less
well defined. There is no {\it a priori} reason to believe that the
source morphology is simple.  However, the complex source geometries
evident for the mass models that produce poor fits likely arise from
the code's attempt to compensate for the poor fits by introducing
complex structure in the source plane.

Using the reconstructed PIEP source model, the source flux was calculated
to be $6.5 \times 10^{-17}$ erg s$^{-1}$ cm$^{-2}$. The overall
magnification of the source is 20, hence this is roughly twice the
value expected from the Ly${\alpha}$ flux alone \citep{Warren1996MN}.
The discrepancy is probably due to continuum around the emission line
which cannot be measured with these observations.
Hence, in the following we assume exactly 1/2 of the source flux is contained in the emission line.
The source emission line luminosity
$L_{Ly{\alpha}}$
is then $4.5 \times 10^{42}$ erg s$^{-1}$ $h_{65}^{-2}$.
Assuming the Ly${\alpha}$ line is not
generated by AGN activity, we estimate the number of O/B stars
\begin{equation}
N_{stars} = \frac{L_{Ly{\alpha}}}{E_{Ly{\alpha}} Q_0 \frac{\alpha^{eff}_{Ly{\alpha}}}{\alpha_B}}
\end{equation}
required to produce the observed luminosity where
$E_{Ly{\alpha}}$ is the energy of a Ly${\alpha}$ photon,
$Q_0$ is the ionising photon luminosity of a single O/B star
from \citet{1996ApJ...460..914V},
$\alpha^{eff}_{Ly{\alpha}} = \alpha_{H_{\beta}} \times (j_{Ly_{\alpha}}/j_{H_{\beta}})(E_{Ly_{\alpha}}/E_{H_{\beta}})$ 
and the ratio
$\frac{\alpha^{eff}_{Ly{\alpha}}}{\alpha_B}$
is calculated to be $8.2$ from tables 2.1 and 4.1 in \citet{1989agna.book.....O}
assuming a temperature of $10^4$K,
although the ratio depends only weakly on temperature.
$Q_0$ differs substantially depending on stellar type, ranging from $10^{49.53}$ for an O5 V star
to $10^{47.9}$ for a B0.5 V star. Using these values, we estimate $N_{stars}$
to be between $10^3$ (all O5 V stars) and $4 \times 10^4$ (all B0.5 V stars).

\begin{table*}
\centering{
\begin{tabular}{lccccl}
Lens & Model & & Image/Source & Area of& Best fitting \\
Model &  $\nu\pm\sqrt{2\nu}$ & $\chi^2$ & pixel size ratio & caustic (arcsec$^2$) & parameters\\ \hline \hline
1. PIEP &   719$\pm$38 & 713.0 & 2 & 0.060 & b=$1.170 \scmd \pm 0.004 $ \\ 
& & & & & $q = 0.917 \pm 0.004 $ \\
& & & & & $\theta_0= 66.2^{\circ} \pm 0.6 $ \\
2. SIS+$\gamma$ & 719$\pm$38 & 876.5 & 2 & 0.050 & b=$1.166 \scmd$ \\
& & & & & $\gamma = 0.077$ \\
& & & & & $\theta_0= 60.4^{\circ}$ \\
3. SIE &  719$\pm$38 & 712.0 & 2 & 0.060 & $b=1.173 \scmd \pm 0.003$ \\
& & & & & $q = 0.775 \pm 0.005 $ \\
& & & & & $\theta_0= 65.9^{\circ} \pm 0.5$ \\
4. SPEMD & 718$\pm$38 & 707.3 & 2 & 0.060 & $b=1.151 \scmd \pm 0.007$ \\
& & & & & $q = 0.76 \pm 0.006 $ \\
& & & & & $\theta_0= 65.9^{\circ} \pm 0.9 $ \\
& & & & & $\beta = 1.08 \pm 0.03 $ \\
5. Constant M/L & 721$\pm$38 & 799.2 & 1.5 & 0.084 & $\kappa_{1/2} = 0.453$ \\
6a. NFW(25\kpc) & 719$\pm$38 & 863.1 & 3 & 0.020 & $\rho_c = 0.3225$ \\
& & & & & $q = 0.95$ \\
& & & & & $\theta_0= 64.4$ \\
6b. NFW(5\kpc) & 719$\pm$38 & 721.2 & 2 & 0.060 & $\rho_c = 1.133 \pm 0.003 $ \\
& & & & & $q = 0.918 \pm 0.003 $ \\
& & & & & $\theta_0 = 66.4 \pm 0.8$ \\
\end{tabular}
\caption{Results of modelling with six mass models. Models which provide an acceptable fit are shown
with $1 \sigma$ errors. Models which do not provide an acceptable fit are shown with the `best fit'
parameters only. The scale factor $b$ for the SPEMD is the same as the lens Einstein radius only
when $\beta = 1$.}
\label{bigtable}
}
\end{table*}

\section{Discussion}
KT03 have undertaken an independent study of the \ER \ system and
before discussing the implications of our results we consider the main
results of the two studies.  KT03 did not find acceptable fits with an
SIE+$\gamma$ model and a single component source.
Given that they used the lensing data only to measure the total mass interior to the images, 
KT03 did not pursue more complex models.
The model of KT03 required significant
external shear and a larger Einstein radius, and consequently their
critical curve was larger and flatter and a larger mass was enclosed
within it.  However, the mass enclosed within a circular
aperture of radius $1.17 \scmd$ is $3.05 h_{65}^{-1} \times 10^{11} \msun$ for the
KT03 model, which agrees well with our models.  This result implies
that the power-law slope of the mass density published in KT03
($\gamma^{\prime}$) can be revised upwards by 0.1 at the radius we
are considering, leading to the best
estimate of the slope to be the isothermal value $\gamma^{\prime}=2.0$
(L. Koopmans, \emph{private communication}).  A simple 1--dimensional
calculation of the expected velocity dispersion for a critical radius
of 1.17\arcsec yields $\sigma = 234\kms$ also in good agreement with
the measurement of the corrected central velocity dispersion of $229
\pm 15$ kms$^{-1}$ from KT03.

Our investigation has identified four mass models [PIEP, SIE, SPEMD and NFW(5\kpc)]
which can reproduce the data.
While the NFW($5\kpc$)
model was successful, the inability of a NFW profile with a more
realistic scale length (25\kpc) to reproduce the properties of the \ER \
system is almost certainly due to the neglect of the baryonic
component in the deflector galaxy. The baryonic component is expected
to contribute significantly to the total surface mass density within
the Einstein radius and the success of the NFW($5\kpc$) model is due
to the ability of the smaller, much higher density, core to mimic the
effect of a significant baryonic contribution at the galaxy centre.
It is quite conceivable that inclusion of the baryonic component, and allowance for it's effects on the dark-matter profile \citep[e.g.][]{2002ApJ...575...87T}, could reconcile the data with the NFW profiles predicted by simulations.

Our modelling has shown that the data alone are able to
tightly constrain the slope of the total mass profile of the galaxy.
While the elliptical isothermal-like models are all consistent with the
data, our models favour a mass profile which is slightly steeper
than isothermal ($\Sigma \propto R^{-1.08 \pm 0.03}$) around the Einstein
radius. The difference in slope corresponds to a $\sim 8\%$ systematic
reduction in the value of $H_0$ which would be inferred from a time-delay
measurement if a pure isothermal model were used.

\begin{figure*}
\centering
\mbox{\subfigure{\psfig{figure=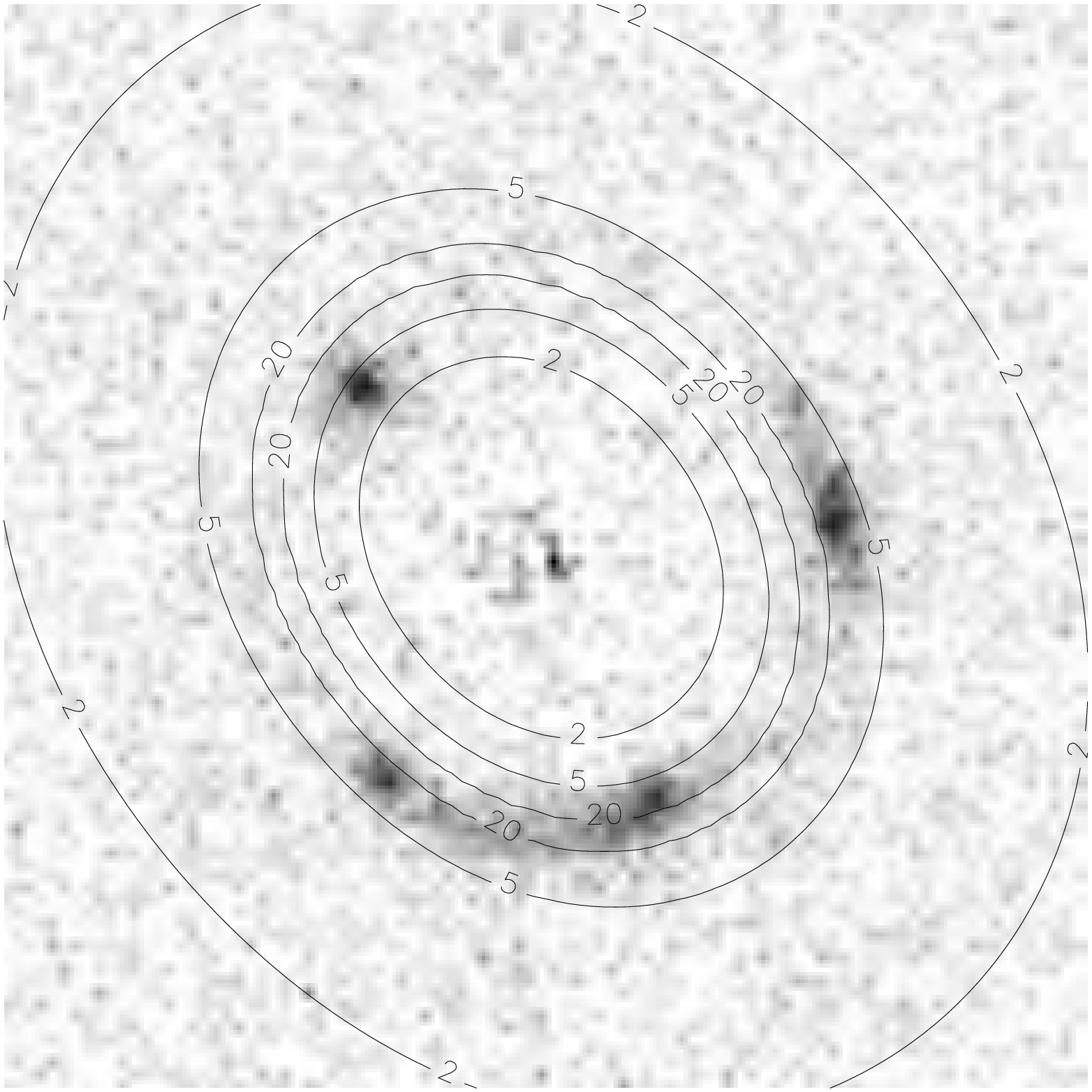,width=55mm}} \quad
      \subfigure{\psfig{figure=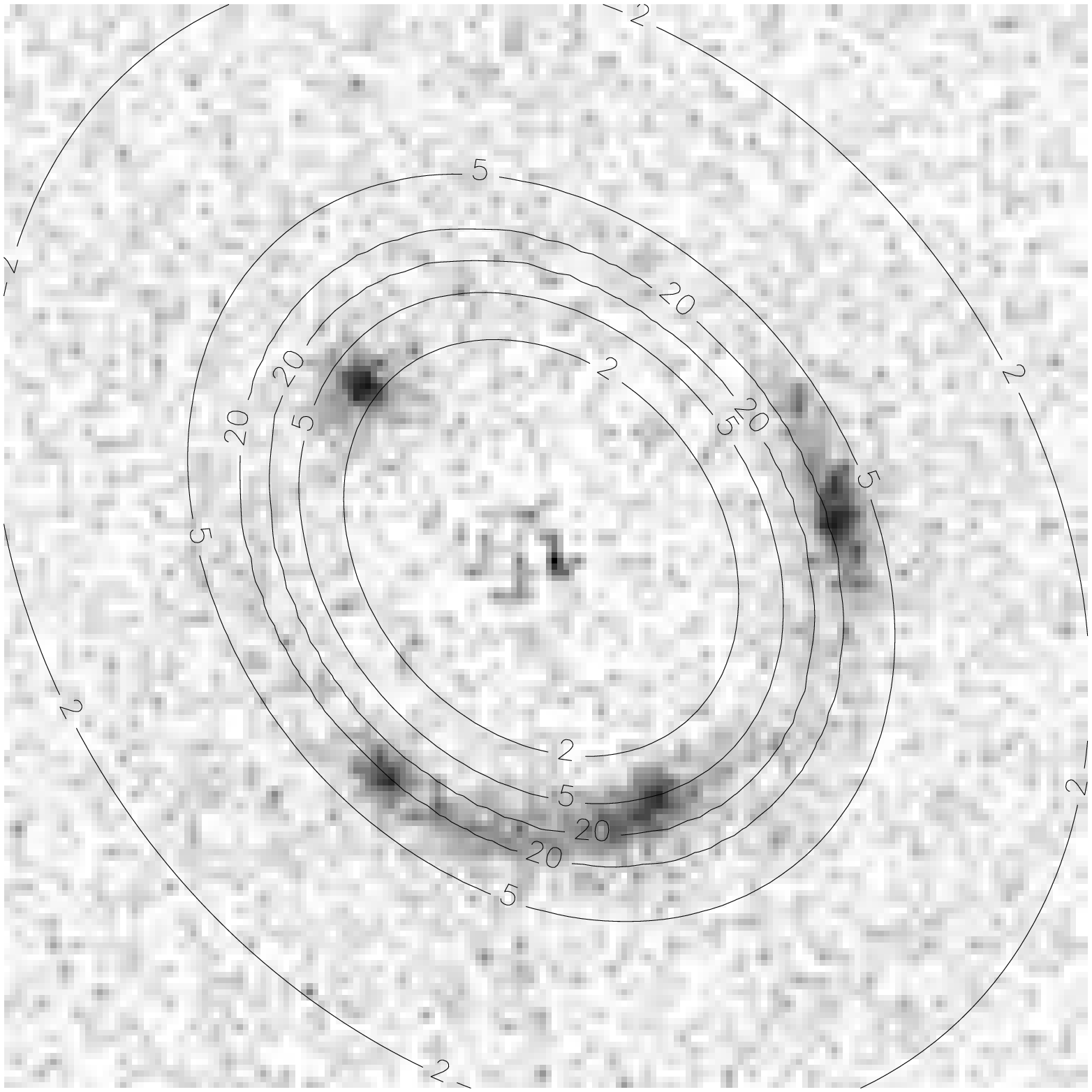,width=55mm}} \quad
      \subfigure{\psfig{figure=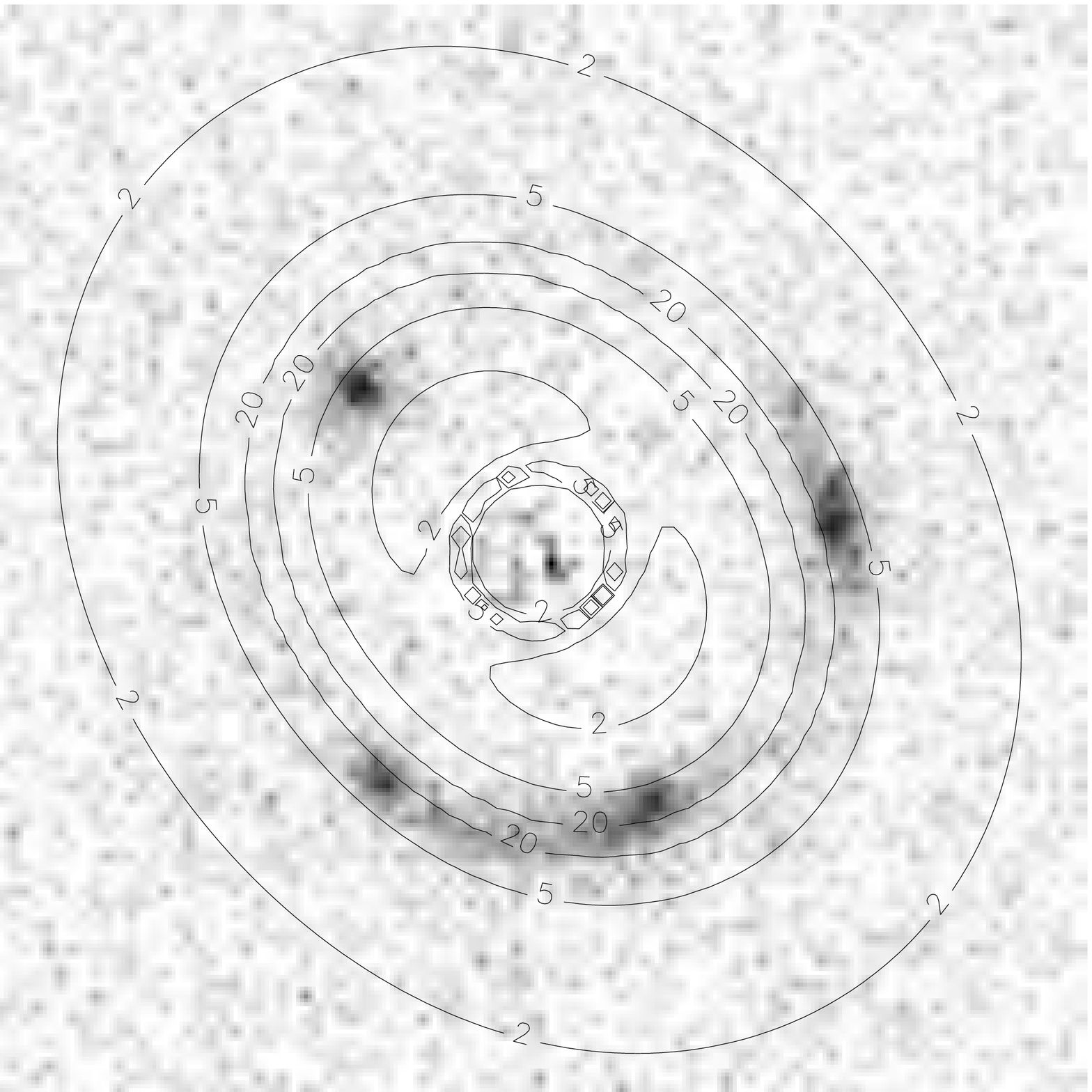,width=55mm}}
}
\caption{Contours of equal magnification for the successful lens
models plotted over the data. From left to right: the PIEP, SPEMD and NFW(5\kpc)
models. Contours show the regions of the image where the magnification
is 2, 5 and 20. The critical line for the lens model is between the
two contours for magnification 20 in each case. The successful models
have remarkably similar critical curves and magnifications around the
locations of the images.}
\label{fig:img_critlines_good}
\end{figure*}

\begin{figure*}
\centering
\mbox{\subfigure{\psfig{figure=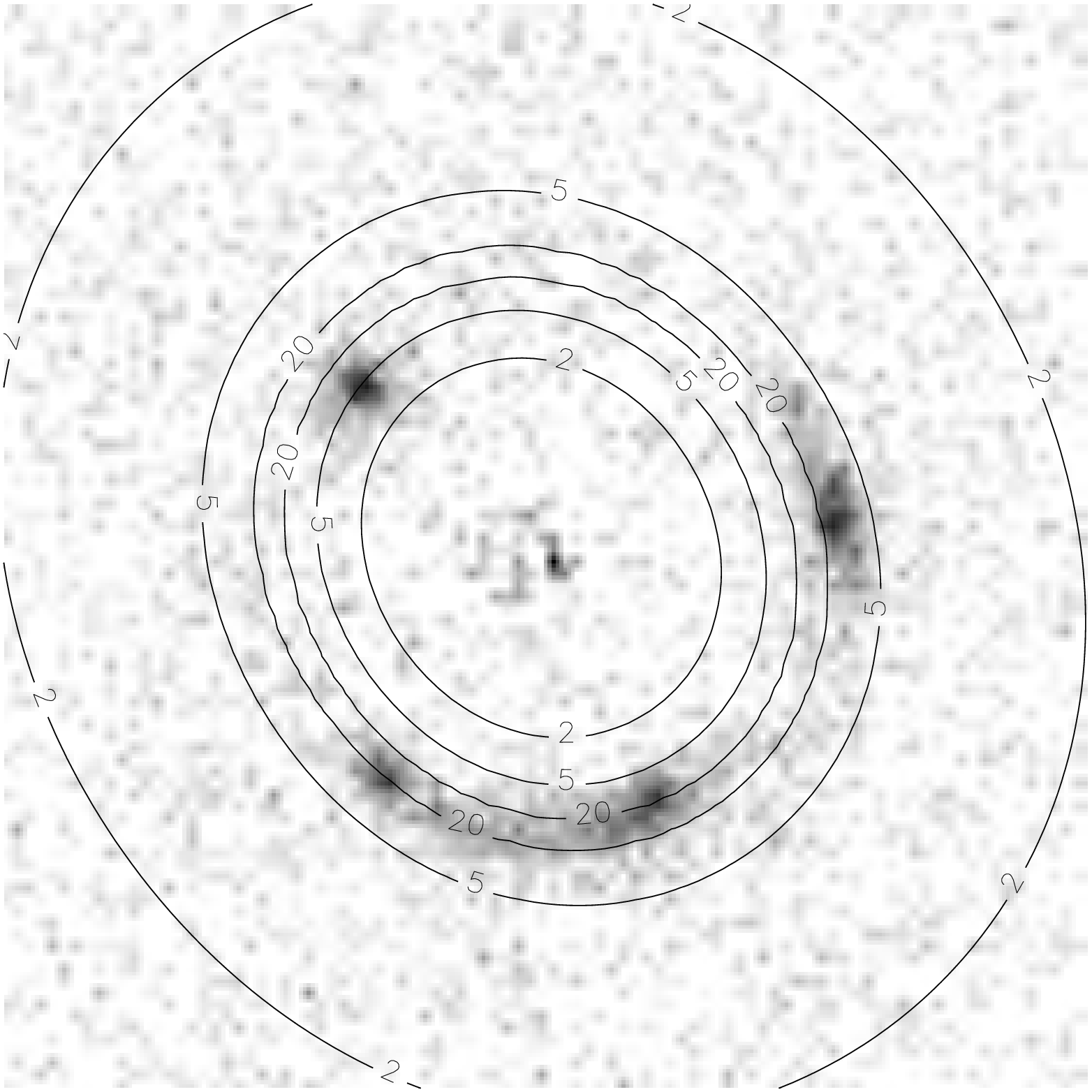,width=55mm}} \quad
      \subfigure{\psfig{figure=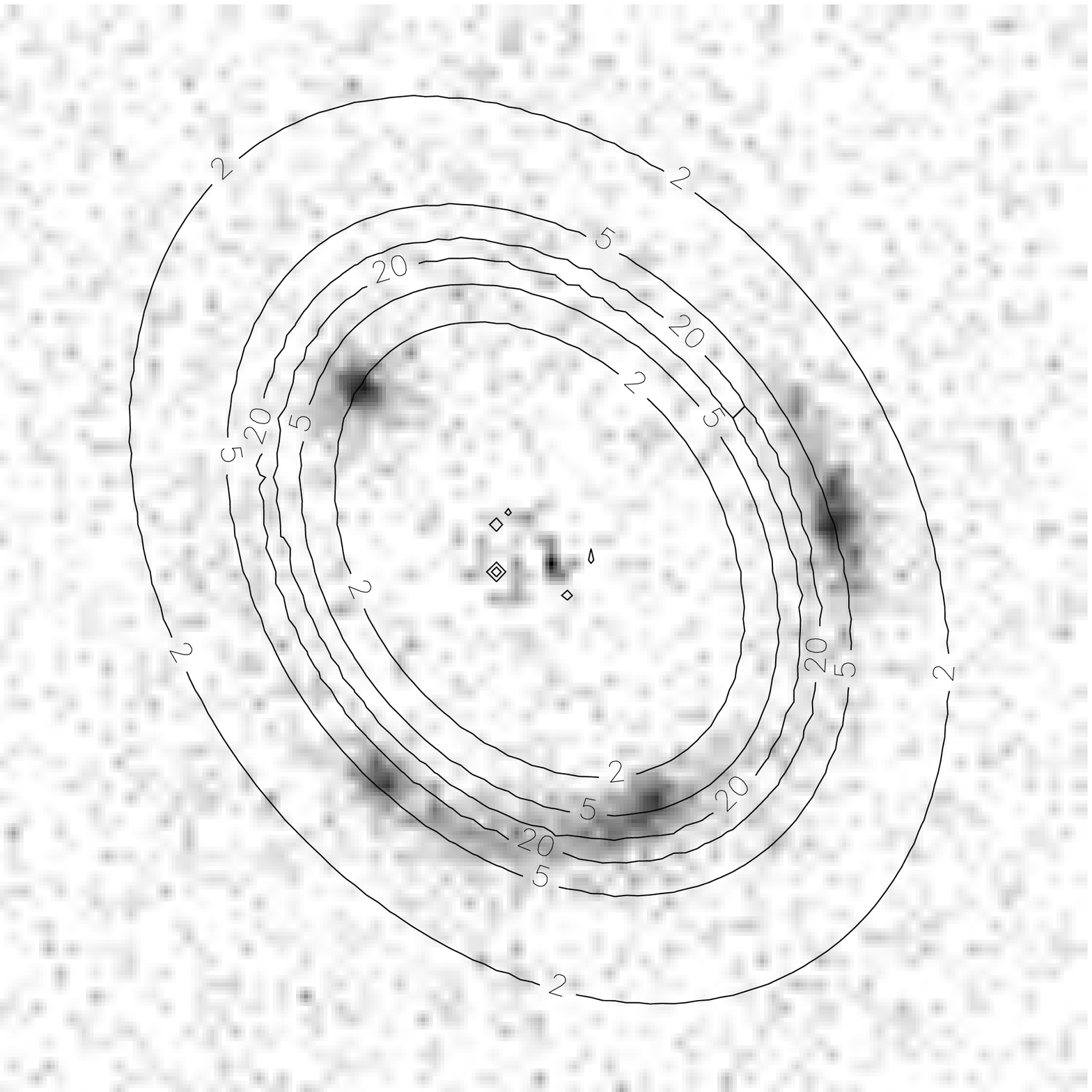,width=55mm}} \quad
      \subfigure{\psfig{figure=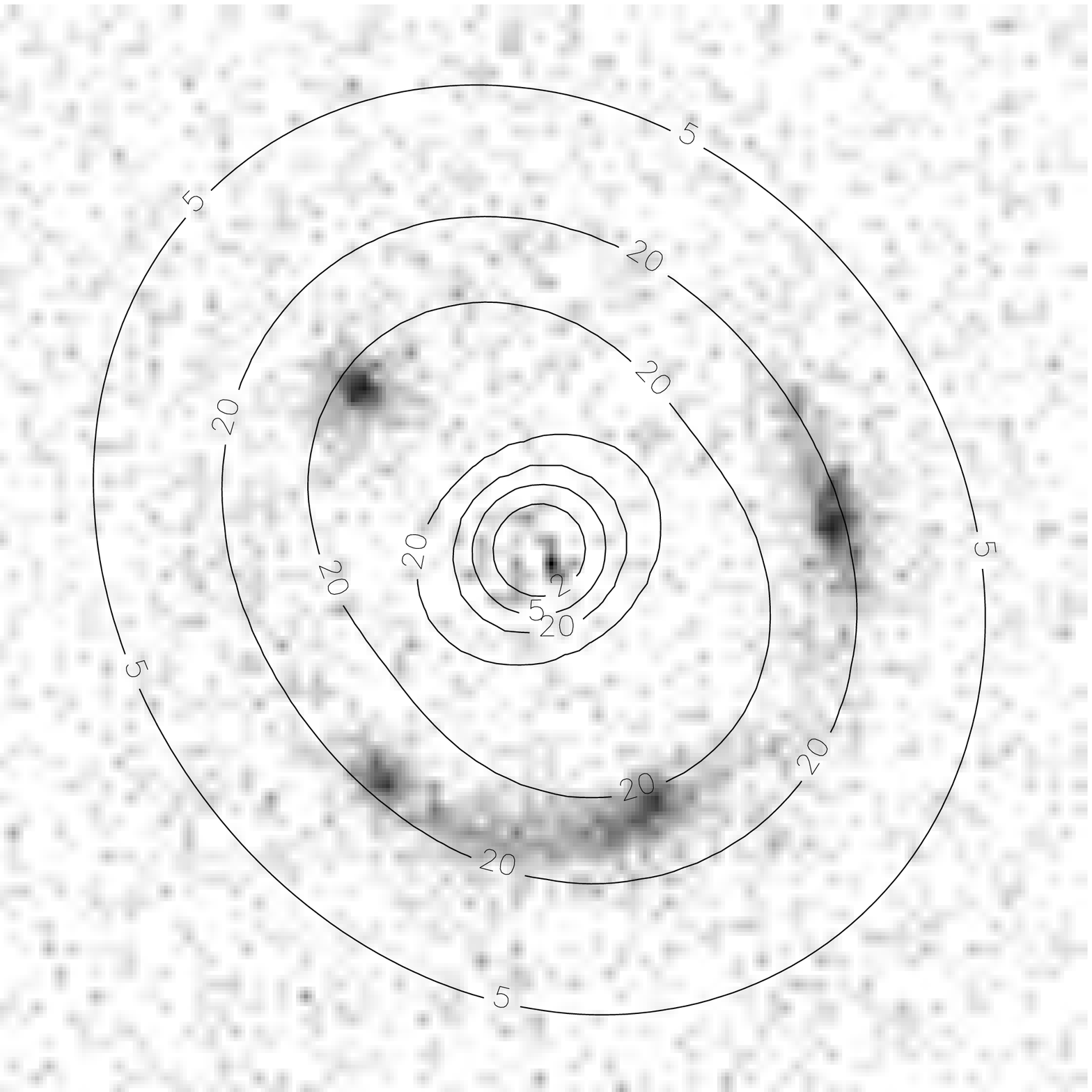,width=55mm}}
}
\caption{Contours of equal magnification for the unsuccessful lens
models plotted over the data. From left to right: the SIS$+\gamma$, constant M/L  and
NFW(25\kpc) models. Contours show the regions of the image where the
magnification is 2, 5 and 20. The outer magnification 2 contour for the
NFW(25kpc) is outside the image region.
The critical line for the lens model is between the two contours for
magnification of 20 in each case.}
\label{fig:img_critlines_bad}
\end{figure*}

Three mass models [SIS$+\gamma$, Constant M/L and NFW(25\kpc)] are not capable
of reproducing the data.  To understand the potential discriminating
power of the data we have plotted contours of equal magnification over
the data for the successful models (Fig.
\ref{fig:img_critlines_good}), and unsuccessful models (Fig.
\ref{fig:img_critlines_bad}).  The diagrams show the magnification in
each region of the image. The critical lines (not shown) lie between
the closely spaced contours where the magnification is equal to a
factor of 20.

Fig. \ref{fig:img_critlines_good} shows the subtle differences between
the successful models.  The SIE model is indistinguishable from the
PIEP model at this image resolution and is not shown.  The SPEMD model
has a slightly more elliptical critical line compared to the PIEP.
In contrast, the NFW(5\kpc) model has a slightly rounder tangential critical line
and some interesting properties near the centre of the lens, with the
presence of a radial critical line before the highly demagnified centre of the lens.
In spite of this property, the
NFW(5\kpc) lens model does not produce any images near the centre of
the lens (given our source position). It is also apparent from Fig.
\ref{fig:img_critlines_good} that the magnification of each of the
images is virtually the same for the successful models.  Images A, B
\& C are the most highly magnified with a magnification factor $\sim
10$, while image D has a lower magnification of $\sim 4$.
Both the similarity in the shape of the magnification contours and of the image
magnifications themselves suggest that lens
models which have elliptical isothermal--like mass distributions on
the scale of the Einstein radius will be able to fit the data (in terms of pure $\chi^2$).

Fig. \ref{fig:img_critlines_bad} shows that the magnification contours
and critical line shape for the unsuccessful models differ
significantly from the cases of the PIEP and SIE models.  Close
examination of the model images show they incorrectly reproduce the
brightness and/or location of the bright regions in the data.  The
SIS$+\gamma$ model produces rounder contours and images B \& C are
closer to the critical line.  This explains why the model does not
reproduce the smaller component of the source well, with the
consequence that images A1, B \& C in the model image are fainter than
the data and image C is also incorrectly located.  The constant M/L
model has magnification contours which are more elliptical than for
the PIEP and SIE models.  All of the images are further from the
critical line and image D in particular is in a region of low ($\sim
2$) magnification, resulting in an image which is the wrong shape
(image B is incorrectly located) and considerably fainter than in the
data.  The NFW(25\kpc) model has very unusual magnification contours and
all of the images are in a high ($\geq 15$) magnification region. This
is a consequence of the small area of the caustic in the model, which
causes the images to be smeared over several pixels;
images B \& D are too large while image D is stretched radially more
than in the data. Image C is not reproduced as a bright region at all.

Thus, the data are sensitive to both the shape (elliptical vs quadrupole) of the mass distribution and the radial mass density profile. To compare the properties of each model, $\kappa$ and $\beta$ (the local logarithmic slope of the surface mass distribution) at each of the image locations (A,B,C,D) were calculated for each lens model.
\begin{table*}
\centering{
\begin{tabular}{lcccc|cccc}
Model & $\kappa_A$ & $\kappa_B$ & $\kappa_C$ & $\kappa_D$ & $\beta_A$ & $\beta_B$ & $\beta_C$ & $\beta_D$  \\ \hline
PIEP  & 0.43 & 0.57 & 0.44 & 0.66 & 1.01 & 1.00 & 1.00 & 1.00 \\
SIE   & 0.43 & 0.57 & 0.44 & 0.67 & 1    & 1    & 1    & 1 \\
SPEMD & 0.41 & 0.56 & 0.42 & 0.67 & 1.08 & 1.08 & 1.08 & 1.08 \\
NFW(5\kpc)   & 0.45 & 0.49 & 0.49 & 0.60 & 1.34 & 1.36 & 1.31 & 1.30 \\
NFW25 & 0.68 & 0.73 & 0.72 & 0.81 & 0.74 & 0.72 & 0.72 & 0.67 \\
Constant M/L & 0.21 & 0.43 & 0.21 & 0.59 & 2.19 & 1.94 & 2.18 & 1.85 \\
SIS + $\gamma$&0.45 & 0.50 & 0.49 & 0.58 &    1 &    1 &    1 & 1
\end{tabular}
\caption{Comparison of $\kappa$ and the local logarithmic slope ($\Sigma \propto R^{-\beta}$) at each image position. The values for images A and D are somewhat uncertain because the images are extended.}
\label{tab:kap_beta_allmodels}
}
\end{table*}
The results are shown in Table \ref{tab:kap_beta_allmodels}. The table shows that: 1) the properties of the PIEP, SIE and SPEMD  are all very similar as expected, and 2) the particular value of $\kappa$ at the image locations is irrelevant, also as expected. The NFW(5\kpc) model has a local slope which is steeper than the PIEP/SIE and SPEMD models. Hence it has generated a poorer $\chi^2$ compared to these models (although still formally acceptable). This is consistent with the data having the ability to distinguish between models based partially upon the local  slope of the mass distribution.

This power in the data to distinguish between models raises the question: if the local slope and ellipticity of the mass distribution are important, what exactly is `local'? In the context of \ER, could we (for example) take the mass inside 0.5\arcsec\ in the SIE model and replace it with an equivalent point mass with no change in the lensing properties?
To quantify this `local' property, the elliptical constant density slab mass models of \citet{1994A&A...284...44S} are used. Consider the mass distribution in the lens as a stack of $N$ constant density elliptical slabs with some elliptical radius $r_e(i)$ [$r_e(i-1) < r_e(i)$ for a monotonically decreasing surface mass density].
Each slab has well defined lensing properties and the deflection angle at any point is simply the sum of contributions from all slabs. For a point inside a slab, the deflection is simply that of a homeoidal ellipse. For a point outside the slab, the deflection is equivalent to a point inside a different slab with foci which are at the same point as the original slab, a `confocal ellipse'.
If a mass distribution is to be distinguishable based on ellipticity and/or radial mass distribution, then the contribution from a slab must not be equivalent to some other slab or point mass.
\citet{1994A&A...284...44S} gives us the answer for \ER.
We define the `effective ellipticity' ($\epsilon_e$) at a point outside an elliptical slab as the ellipticity of an equivalent confocal elliptical slab. It is easy to show that $\epsilon_e \propto (r_e/r)^2$ for $r > r_e$. That is, the effective ellipticity decreases rapidly for points increasingly further away from the outside of the slab. For $r \gg r_e$, the slab looks like a point mass, as expected. Hence, for the models presented here where the error on ellipticity (axis ratio) is $\sim 10\%$, the mass inside $\sim 0.03$\arcsec\ can be replaced with a point mass with no significant change in the lensing properties around the images. In a similar way, slabs with progressively larger radii (larger than the image radius) have progressively lower surface density ($\kappa \propto r^{-1}$ for an SIE), hence their contribution to the overall ellipticity is $\propto r^{-1}$. Again, with an error of $\sim 10\%$ on the measured ellipticity, slabs with a radius $\gtrsim 10 \arcsec$ are indistinguishable from a single mass sheet.
Using this simple argument, we have defined what `local' means for \ER. The implication is that this lens cannot provide information about the mass distribution in the very centre of the galaxy or in the halo very far from the galaxy but there is a substantial range in between which does contribute to the image configuration.

A final noteworthy property of \ER\ is the lack of necessity of any
external shear for a successful lens model. In general, there is likely to be a shear
contribution due to the environment. However, the lens is an isolated
field elliptical with nearest neighbour $\sim 1^{\prime}$ away \citep{Warren1999}.
The shear generated by this object will be small ($\gamma \sim 0.02$)
so it is not surprising that no shear was needed to model this system.
The large shear required in the model of KT03 was a consequence of having
an oversimplified source. We emphasise the importance of using a suitable
source model for resolved lenses before conclusions are made about the lens
mass distribution.

The combination of this work with that of KT03 has
strongly constrained the mass distribution in the lens galaxy, showing
it to be very close to isothermal (at least in the region of the images).
Well constrained lens models are required for $H_0$
measurements using lensing time delays, thus \ER\ is an excellent
candidate for monitoring of supernova events in the source.
The lack of external shear also reveals a good deal about
the overall mass distribution in the lensing galaxy.
\citet{1998ApJ...509..561K} and 
\citet{Kochanek2002sgdh.conf...62K} have studied the alignment between
visible matter and total matter distributions in lens galaxies and
found they have an RMS misalignment $< 10^{\circ}$.
Our result adds weight to their conclusions.

\section{Conclusions}\label{conclusions}
We have performed a detailed analysis of \ER\ using data from WFPC2 on
board HST. The data show that the lensed image has four distinct
bright regions and two regions of extended emission. We have modelled
the image with six different lens models using sophisticated
software based on the \textsc{LensMEM} algorithm. The software generates a
best-fitting image and reconstructs the source brightness profile
using a non-parametric source model.

We have tested the ability of the data to distinguish between a
variety of lens models including isothermal, power-law, constant M/L
and NFW mass distributions.  We find that the `canonical' SIE and
PIEP lens models fit the data well and that a power-law model ($\Sigma
\propto R^{-\beta}$) favours $\beta = 1.08 \pm 0.03$,
slightly steeper than isothermal.  In addition we find
that we can fit the data with a mass profile based on the NFW profile,
but only when the scale length is too small and central density too
high for a dark matter halo expected in a large elliptical galaxy.
Conversely, we find that a mass model based on constant M/L or a
realistically-sized NFW halo cannot fit the data.  We also find that
the simple SIS$+\gamma$ lens model does not fit the data despite being
an isothermal model.  The data are good enough to distinguish between
elliptical lens models and the SIS$+\gamma$ which is a first-order
approximation of an elliptical model.  The difference appears to be in
the shape of the critical curve between the elliptical models and
external shear model.  We have discovered also that the source is
actually a double. This has caused problems for models which assume
the bright regions in the image come from a single source.

Our results are somewhat different to those of KT03 who found a larger Einstein
Radius by $15\%$. However, their model assumed a single source and
included a substantial shear component.  The mass enclosed by KT03's
model within a $1.17 \scmd$ aperture is consistent with our model.
Our models show that the image locations and brightnesses can be
explained neatly with the simple SIE/PIEP mass models and the double
source. We calculate the total M/L$_B$, inside 1.17\scmd,
to be $4.7 h_{65} \pm0.3$ compared to $5.4\pm0.8$, at a larger radius of 1.34\scmd, in KT03.

We have qualitatively evaluated the behaviour of the lens models
around the location of the lensed images by plotting contours of
constant magnification. We found that the SIE and PIEP models are
indistinguishable for these data and that any elliptical mass
distribution with an isothermal-like density profile around the images
is likely to be able to fit the data. We found that the lens models
which could not fit the data produce images with incorrect
magnification and/or location.

Finally, we note that the lens model requires no external shear to explain the data. The well constrained mass distribution, lack of
external shear and isolation of the lens galaxy all suggest that \ER\
is an ideal candidate for further work such as monitoring for
supernova events in the source, or making a detailed study of
the properties of the galaxy's dark halo.

\newcommand{\nat}{Nat}
\newcommand{\mnras}{MNRAS}
\newcommand{\aj}{AJ}
\newcommand{\pasp}{PASP}
\newcommand{\aap}{A\&A}
\newcommand{\apjl}{ApJ}

\end{document}